\documentclass{pasj00}
\newcommand{\Msolar}{\mbox{\,$M_{\odot}$}}        % solar mass
\newcommand{\Lsolar}{\mbox{\,$L_{\odot}$}}        % solar luminosity

\begin{document}
\SetRunningHead{W.Bian and Y.Zhao}{Black Hole -- Bulge Relation
for Narrow-Line Objects}
\title{Black Hole -- Bulge Relation for Narrow-Line Objects}
\author
{Wei-Hao B{\small IAN}\altaffilmark{1,2} and Yong-Heng Z{\small
HAO} \altaffilmark{1}} \altaffiltext{1}{National Astronomical
Observatories, Chinese Academy of Sciences, Beijing 100012, China}
\altaffiltext{2}{Department of Physics, Nanjing Normal University,
Nanjing 210097, China } \email{whbian@lamost.bao.ac.cn}

\Received{2002 October 23} \Accepted{2002 November 20}

\maketitle
\begin{abstract}
It has been thought that narrow-line Seyfert 1 galaxies are likely
to be in the early stages of the evolution of active galaxies. To
test this suggestion, the ratios of the central massive black hole
(MBH) mass to the bulge mass ($M_{\rm bh}/M_{\rm bulge}$) were
estimated for 22 Narrow Line AGNs (NL AGNs). It is found that NL
AGNs appear to have genuinely lower MBH/Bulge mass ratio ($M_{\rm
bh}/M_{\rm bulge}$). The mean $ log(M_{\rm bh}/M_{\rm bulge})$ for
22 NL AGNs is $-3.9\pm 0.07$, which is an order of magnitude lower
than that for Broad Line AGNs and quiescent galaxies. We suggest a
nonlinear MBH/Bulge relation and find there exists a relation
between the $M_{\rm bh}/M_{\rm bulge}$ and the velocity
dispersion, $\sigma$, derived from the [O ${\rm \small III}$]
width. A scenario of MBH growth for NL AGNs is one of our
interpretations of the nonlinear MBH/Bulge relation. The MBH
growth timescales for 22 NL AGNs were calculated, with a mean
value $(1.29\pm0.24)\times 10^{8}$ yr. Another plausible
interpretation is also possible: that NL AGNs occur in low-$M_{\rm
bulge}$ galaxies and that in such galaxies $M_{\rm bh}/M_{\rm
bulge}$ is lower than that in galaxies with a higher $M_{\rm
bulge}$, if we consider that NL AGNs already have their ``final''
$M_{\rm bh}/M_{\rm bulge}$. More information of the bulge in NL
AGNs is needed to clarify the black hole -- bulge relation.
\end{abstract}

\section{Introduction}

Evidence shows that there is a strong connection between active
galactic nuclei and their host galaxies. Within the framework of
the hierarchical dark-matter cosmology, the formation and
evolution of galaxies and their active nuclei is intimately
related (Fabian 1999; Haehnelt et al. 1998; Mathur 2000).

Magorrian (1998) found that the central massive black hole (MBH)
mass is proportional to the mass of the host bulge in a sample of
nearby galaxies, hereafter referred to as the MBH/Bulge relation,
with the MBH mass being about 0.006 of the bulge mass. Laor (1998)
also found this relation for 14 bright quasars. Recent research
using higher quality HST data and a more careful treatment of the
modelling uncertainties give lower values of the central MBH
masses in nearby galaxies, with an average MBH-to-bulge mass ratio
of about 0.001 and a nearly linear MBH -- Bulge relation, $M_{\rm
bh}\propto L^{1.08}_{\rm bulge}$ (Merritt, Ferrarese 2001;
Kormendy, Gebhardt 2001; McLure, Dunlop 2001, 2002).

However, Laor (2001) suggested a nonlinear MBH/Bulge mass
relation, $M_{\rm bh}\sim M^{1.54}_{\rm bulge}$, and showed that
the mean MBH-to-bulge mass ratio is therefore not a universal
constant, which is related to the bulge masses. The bulge
information of late-type spirals and of narrow line Seyfert 1
galaxies (NLS1s) (both predicted to have low $M_{\rm bh}$) can
test the MBH/Bulge nonlinear relation. Wandel (2002) revised the
MBH/Bulge relation for a sample of 55 AGNs and 35 inactive
galaxies. He found that broad-line AGNs (BL AGNs) have an average
MBH/Bulge mass ratio of $\sim$ 0.0015 and strong correlations,
$M_{\rm bh}\propto L^{0.9}_{\rm bulge} $, $M_{\rm bh}\propto
\sigma ^{3.5-5}$. For a few narrow-line AGNs (NL AGNs) in Wandel
(2002), the average MBH/Bulge mass ratio is really lower,$\sim
10^{-4}$. Mathur (2000) has proposed that the NLS1s are likely to
be active galaxies in an early stage of evolution, and therefore
have a lower MBH/Bulge mass ratio than BL AGNs. Mathur et al.
(2001) have estimated the central MBH mass for 15 NLS1s by fitting
their spectral energy distributions with the accretion disk and
corona model (Kuraskiewicz et al. 2000), and found the mean mass
ratio of the MBH/Bulge to be 0.00005, lower by a factor of 30
compared to that for broad-line AGNs.

It is found that the central MBH mass is not only related to the
bulge luminosity, but also to the bulge velocity dispersion.
Ferrarese and Merritt (2000) and Gebhardt et al.(2000a) have found
that the MBH mass of inactive galaxies is better correlated with
the stellar velocity dispersion in the host bulge than with the
bulge luminosity, and that the relations are respectively $ M_{\rm
bh}\propto \sigma ^{4.80}$ and $M_{\rm bh}\propto \sigma ^{3.75}$.
Gebhardt et al. (2000b) and Ferrarese et al. (2001) showed that
the MBH masses of a few Seyfert galaxies from reverberation
mapping are consistent with the relation between the MBH mass and
galaxy velocity dispersion which they have found in inactive
galaxies.

The theoretical interpretation for the MBH/Bulge relation is
discussed based on several models. One model is about
merger-driven starbursts with black hole accretion (Kauffmann,
Haehnelt 2000). Some models are based on black hole accretion
influencing the star formation and gas dynamics in the host
galaxies (Silk, Rees 1998; Blandford 1999). Adams et al. (2001)
presented an idealized model of the collapse of the inner part of
protogalaxies, and assumed the MBH mass is determined when the
centrifugal radius of the collapse flow exceeds the capture radius
of the central MBH. They produced the observed relation between
the MBH mass and the galactic velocity dispersion, and predicted
the mass ratio of the MBH/Bulge: $M_{\rm bh}/M_{\rm bulge} \approx
0.004(\sigma/200~ km~ s^{-1})$. Wang et al. (2000) presented a
model which could explain the MBH/Bulge relation in AGNs and the
dependence on the environmental parameters of the host galaxies,
such as the gas or stellar velocity dispersion, as well as the
relation of the central star burst and accretion process during
galactic interaction. They also discussed the mass ratio of
MBH/Bulge based on a unified formation scheme, where the bulge
formation and nucleus activity are triggered  by galaxy mergers or
tidal interactions, and found a correlation of the mass ratio of
the MBH/Bulge to be roughly $M_{\rm bh}/M_{\rm bulge} \propto
\sigma^{1.4}$.

It is important to investigate the lower limit of the MBH/Bulge
mass ratio because it will reveal physical links between the bulge
and the MBH. There has been a progress concerning the estimation
method of the central MBH mass in AGNs (Wandel et al. 1999; Ho
1999; Kaspi 2000; Wang, Lu 2001). NLS1s are suggested to have
smaller central MBH masses with higher accretion rates close to
the Eddington limit. Therefore, NLs1s could play a particular role
to understand the formation of the bulge and central MBH in
galaxies.

In this paper, we investigate the MBH/Bulge relation in NL AGNs
compared with BL AGNs using the recent estimation of MBH masses of
NL AGNs. In section 2 we present a sample of the NL AGNs, along
with the estimated mass of the central MBH and the bulge. In
section 3 we explore the MBH/Bulge relation for the BL AGNs and NL
AGNs. The result and a discussion are presented in section 4, and
in section 5 we summarize our conclusion. All of the cosmological
calculations in this paper assume $H_{0}=75~ km~ s^ {-1}$, $\Omega
=1.0$, $\Lambda=0$.

\section{The Sample of Narrow Line AGNs}

In order to investigate MBH/Bulge relation in NL AGNs, we used
available data of the bulge luminosity (Mackenty 1990; Whittle
1992; Bahcall et al. 1997; Malkan et al. 1998) and central MBH
mass (Veron-Cetty et al. 2001; Wang, Lu 2001) for NL AGNs in the
literature. We selectd NL AGNs with the BMH mass and the bulge
luminosity. Veron-Cetty et al. (2001) have compiled 83 objects
known to us before 1998 January either to be NLS1s or to have a
``broad'' Balmer component narrower than $2000~ km~ s^{-1}$, north
of $\delta=-25^{o}$, bright than $B=17.0$ and with $z<0.1$. The
measurement with a moderate resolution of 3.4$~ \rm{\AA}$ for 59
NLS1s of the instrument-subtracted [O ${\rm \small III}$] and
H$\beta$ widths as well as the optical magnitude at $B$ band are
listed in table 2 and table 3 in Veron-Cetty et al.(2001), which
are used to calculate MBH masses. We obtained the bulge absolute
blue magnitude ($M^{\rm bulge}_{B}$) to calculate the mass of the
bulge. The number of NLS1 suitable for studying the NLS1s
MBH/Bulge relation is limited because there is so little
information about the NLS1s bulge luminosity. We obtained a sample
of 22 NL AGNs (table 1). Wandel (2002) derived the MBH/Bulge
relation for 46 BL AGNs, 9 NL AGNs, and 35 quiescent galaxies. Our
sample includes all 9 NL AGNs in the Wandel (2002) sample.

\subsection{Determination of the MBH Mass}
The central MBH masses of only 6 NL AGNs (Mrk 335, NGC 4051, 3C
120, Mark 110, Mrk 590, PG 1704) in our sample were estimated from
the reverberation mapping method. For the other 16 NL AGNS, we
estimated the size of the broad line region (BLR) using an
empirical correlation between the size and the monochromatic
luminosity at 5100 $\rm{\AA}$ (Kaspi et al. 2000),
\begin{equation}
R_{\rm BLR}=32.9(\frac{\lambda L_{\lambda}(5100~
\rm{\AA})}{10^{44} erg~ s^{-1}})^{0.7} ~~\rm{lt-d}, \label{eqn1}
\end{equation}
\noindent where $\lambda L_{\lambda}(5100~ \rm{\AA)}$ was
estimated from the $B$-magnitude  by adopting an average optical
spectral index of -0.5 and accounting for the galactic redding and
$k$-correction. If the H$\beta$ widths reflect the Keplerian
velocity of the line-emitting BLR material around the central MBH,
then the so-called virial mass estimated for the central MBH  is
given by
\begin{equation}
M_{\rm bh}=R_{\rm BLR} V^{2} G^{-1}, \label{eqn2}
\end{equation}
\noindent where $G$ is the gravitational constant, and $V$ is the
velocity of the line-emitting material. $V$ can be estimated from
the H$\beta$ width. Assuming random orbits, Kaspi (2000) related
the $V$ to the FWHM of the H$\beta$ emission line by
$V=(\sqrt{3}/2) \rm FWHM_{[H\beta]}$. The calculated central MBH
masses for 15 NLS1s are listed in table 1 (Wang, Lu 2001).

\subsection{Determination of the Bulge Mass}
We estimate the bulge masses of the NLS1s from the bulge absolute
blue magnitude ($M^{\rm bulge}_{B}$) of the host galaxies (Laor
1998; Wandel 1999; Mathur 2000). $M^{\rm bulge}_{B}$ was
calculated from the galaxy's total bulge blue magnitude ($M^{\rm
total}_{B}$) by the Simien and de Vaucouleurs (1986) equation,
\begin{equation}
M^{\rm bulge}_{B} = M^{\rm total}_{B} - 0.324~ \tau + 0.054~
\tau^{2} - 0.0047~ \tau^{3}, \label{eqn3}
\end{equation}
\noindent where $\tau=T+5$ and $T$ is the Hubble-type of the
galaxy. We adopted a canonical Hubble-type of Sa for Mrk 734, Mrk
486, and Mrk 1239. For Mrk 1044, we took the host galaxy magnitude
from MacKenty (1990), who included nuclear emission in the total
blue magnitude. Hence, in table 1 we quote the blue magnitude as
an upper limit. We then use the relation between the bulge $B$ and
$V$ magnitudes. We used $B-V=0.8$, and calculated the bulge
luminosity from the empirical formula,
\begin{equation}
{\rm log}(L_{\rm bulge}/\Lsolar)=0.4 (-M^{\rm bulge}_{V} + 4.83).
\label{eqn4}
\end{equation}
\noindent Finally, we used the mass and luminosity relation for
normal galaxies from Magorrian et al. (1998),
\begin{equation}
{\rm log}(M_{\rm bulge}/\Msolar) = 1.18 {\rm log}(L_{\rm
bulge}/\Lsolar) - 1.11. \label{eqn5}
\end{equation}
The calculated bulge masses of NLS1s are listed in table 1.

\begin{table*}
\caption{Central MBH and the bulge properties of 22 NL AGNs.}
\begin{center}
\begin{tabular}{lccccccc}
\hline\hline
Name & log $M_{\rm bh}$ & log $M_{\rm bulge}$ & log $\frac{M_{\rm bulge}}{M_{\rm bh}}$ & log $L_{\rm bol}$ & log $\frac{L_{\rm bol}}{L_{\rm Edd}}$ & [O ${\rm \small III}$] & $t_{\rm s}$ \\
(1)&(2)&(3)&(4)&(5)&(6)&(7)&(8)\\
\hline
Mrk~335        & 6.80  & 10.56  & -3.76 & 44.79 & -0.12  & 245  & 1.67   \\
Mrk~359        & 6.23  & 10.11  & -4.59 & 44.66 &  0.32  & 180  & 0.75   \\
Mrk~705        & 6.92  & 11.11  & -4.20 & 44.79 & -0.24  & 365  & 2.06   \\
Mrk~124        & 7.20  & 10.51  & -3.31 & 45.17 & -0.15  & 380  & 0.51   \\
Mrk~142        & 6.67  & 10.59  & -3.91 & 44.77 & -0.02  & 260  & 0.96   \\
Mrk~42         & 6.00  & ~9.70  & -3.70 & 44.40 &  0.28  & 220  & 0.38   \\
NGC~4051       & 6.11  & 10.05  & -3.94 & 43.56 & -0.66  & 200  & 4.36   \\
Mrk~766        & 6.63  & 10.62  & -3.99 & 44.51 & -0.24  & 220  & 1.73   \\
Akn~564        & 6.46  & 10.62  & -4.16 & 45.04 &  0.47  & 220  & 0.39   \\
Mrk~486        & 7.03  & 10.66  & -3.63 & 45.04 & -0.11  & 400  & 0.85   \\
Mrk~734        & 7.34  & 11.27  & -3.93 & 45.37 & -0.08  & 180  & 1.14   \\
Mrk~1239       & 6.38  & 10.40  & -4.02 & 44.65 &  0.16  & 400  & 0.71   \\
Mrk~382        & 6.61  & 10.82  & -4.21 & 44.78 &  0.43  & 155  & 1.31   \\
Mrk~493$^{\dagger}$  & 6.11  & 10.07  & -3.96 & 44.74 &  0.01  & 315  & 0.81   \\
Mrk~1044$^{\ddagger}$ & 6.23  & 10.76  & -4.53 & 44.52 &  0.29  & 335  & 0.30   \\
3C~120         & 7.36  & 10.72  & -3.36 & 45.34 & -0.13  & 230  & 0.56   \\
Mrk~110        & 6.75  & 10.74  & -3.99 & 44.71 & -0.15  & 290  & 1.42   \\
Mrk~590        & 7.25  & 11.03  & -3.78 & 44.63 & -0.73  & 400  & 4.34   \\
0157+001      & 7.7   & 11.79  & -4.09 & 45.62 & -0.19  &  -   & 1.70   \\
1402+26       & 7.28  & 10.61  & -3.33 & 45.13 & -0.26  &  -   & 0.70   \\
1704+60       & 7.57  & 11.13  & -3.56 & 46.33 &  0.65  & 440  & 0.13   \\
2247+140      & 7.59  & 11.55  & -3.96 & 45.47 & -0.23  & -    & 1.66   \\
\hline
\end{tabular}
\end{center}
$\ast$ Col.1: name, Col.2: log of the estimated MBH mass in
$\Msolar$, Col.3: log of the estimated bulge mass in $\Msolar$,
Col.4: log of the MBH/Bulge mass ratio, Col.5: log of the the
bolometric luminosity in unit of $\rm erg~ s^{-1}$, Col.6: log of
the ratio of the bolometric luminosity to the Eddington
luminosity, Col.7: FWHM (in $km~s^{-1}$) of [O ${\rm \small
III}$],
Col.8: growth timescale for NLS1s in unit of $10^{8}$ yr. \\
$\dagger$
The bulge absolute blue magnitude from Malkan (1998). \\
$\ddagger$ The bulge absolute blue magnitudes are adopted from
MacKenty (1990), the others are adopted from Whittle (1992).
\end{table*}

\section{MBH -- Bulge Relation}
\subsection{$M_{\rm bh}/M_{\rm bulge}$ Distribution}

For 22 NL AGNs we found the mean ${\rm log} (M_{\rm bh}/M_{\rm
bulge})$ to be $-3.9\pm 0.07$, which is an order of magnitude
lower than that of BL AGNs. Mathur et al. (2001) found a smaller
$M_{\rm bh}/M_{\rm bulge}$ value of 0.00005. The difference is due
to their underestimated MBH masses from the spectral fitting.
Wandel (2002) also found a smaller ${\rm log} (M_{\rm bh}/M_{\rm
bulge})$ value for 9 NL AGNs ($-3.9\pm 0.27$), which is consistent
with our results.

\subsection{The Nonlinear $M_{\rm bh}$ -- $M_{\rm bulge}$ Relation}
\begin{figure}
\begin{center}
\FigureFile(80mm,80mm){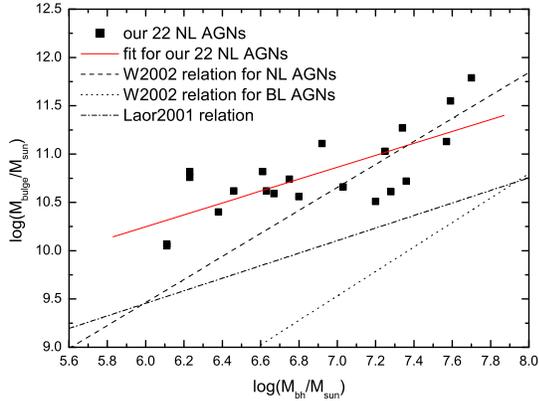}
\end{center}
\caption{Bulge mass versus the MBH mass. The squares denote NL
AGNs . The thin solid line is the best fit to 22 NL AGNs. The
dashed line and dotted line show the MBH/Bulge mass fit line
[Wandel (2002) for NL AGNs and BL AGNs, respectively]. We also
plotted the relation for 40 galaxies in dot-dashed line founded by
Laor (2001). }
\end{figure}

In figure 1 we plot the bulge mass vs. the MBH mass for 22 NL
AGNs. There is a significant correlation with a correlation
coefficient of $R=0.74$, corresponding to a probability of
$P=1.23\times 10^{-4}$ that the correlation is caused by a random
factor. The best linear fit is
\begin{equation}
{\rm log}(M_{\rm bulge}/\Msolar) = (0.62\pm 0.13) {\rm log}(M_{\rm
bh}/\Msolar) + (6.55\pm 0.88).
\end{equation}

In figure 1 we also plot the relations found by Wandel (2002) and
Laor (2001). Our fit line is higher compared to that for BL AGNs
in Wandel (2000). The MBH/Bulge mass relation for our 22 NL AGNs
is consistent to that for 9 NL AGNs in Wandel (2002). The result
of the lower MBH/Bulge mass ratio for NL AGNs is reliable. We
suggested the nonlinear MBH/Bulge relation; namely, the MBH/Bulge
mass ratio is not a constant. In the next section we discuss the
relation between $M_{\rm bh}/M_{\rm bulge}$ and the bulge velocity
dispersion, $\sigma$.

\subsection{$M_{\rm bh}/M_{\rm bulge}$ -- $\sigma$ Relation}
\begin{center}
\begin{table}
\caption{Central MBH and the bulge properties of AGNs (Wandel
2002), except our 22 NL AGNs.}
\begin{center}
\begin{tabular}{lcccc}
\hline \hline
Name      & log $M_{\rm bulge}$ & log $M_{\rm bh}$& log $\frac{M_{\rm bh}}{M_{\rm bulge}}$& [O ${\rm \small III}$]  \\
(1)&(2)&(3)&(4)&(5)\\
\hline
0736+017  &   11.46    &  7.99 &  -3.47        &  540                        \\
0953+41   &   11.08    &  8.39 &  -2.69        &  640                        \\
1004+13   &   11.70    &  9.09 &  -2.61        &  470                        \\
1020-103  &   11.36    &  8.35 &  -3.01        &  430                        \\
1116+215  &   11.51    &  8.22 &  -3.29        &  380                        \\
1202+28   &   11.08    &  8.29 &  -2.79        &  500                        \\
1217+023  &   11.55    &  8.40 &  -3.15        &  380                        \\
1226+02   &   11.84    &  8.61 &  -3.23        &  470                        \\
1302-10   &   11.41    &  8.30 &  -3.11        &  710                        \\
1425+267  &   11.36    &  9.32 &  -2.04        &  310                        \\
1444+40   &   11.13    &  8.06 &  -3.07        &  540                        \\
1545+21   &   11.32    &  8.93 &  -2.39        &  610                        \\
2135-14   &   11.41    &  8.94 &  -2.47        &  360                       \\
2141+175  &   11.55    &  8.74 &  -2.81        &  1120                       \\
Mrk79$^\dagger$      &   10.15  &  8.02    &   -2.13    &       350         \\
Mrk817$^\dagger$     &   10.33  &  7.56    &   -2.77    &       330         \\
NGC3227$^\dagger$    &    9.47  &  7.69    &   -1.78    &       485         \\
NGC3516$^\dagger$    &   10.69  &  7.36    &   -3.33    &       250         \\
NGC4151$^\dagger$    &    9.31  &  7.08    &   -2.23    &       425         \\
NGC4593$^\dagger$    &   10.21  &  6.91    &   -3.30    &       255         \\
NGC5548$^\dagger$    &   10.31  &  7.83    &   -2.48    &       410         \\
NGC6814$^\dagger$    &    9.81  &  7.08    &   -2.73    &       125         \\
\hline
\end{tabular}
\end{center}
$\ast$ Col.1, name; Col.2, log of the estimated bulge mass in
$\Msolar$; Col.3, log of the estimated MBH mass in $\Msolar$;
Col.4, log of the MBH/Bulge mass ratio; Col.5, FWHM (in
$km~s^{-1}$) of [O ${\rm \small III}$]. The MBH and bulge mass are
all adopted from Wandel (2002).
\\ $\dagger$ Seyfert galaxies, the others are PG quasars. The
FWHM of [O ${\rm \small III}$] for PG quasars are adopted from
Marziani et al.(1996) and that for Seyfert galaxies are from
Whittle(1992).
\end{table}
\end{center}

\begin{figure}
\begin{center}
\FigureFile(80mm,80mm){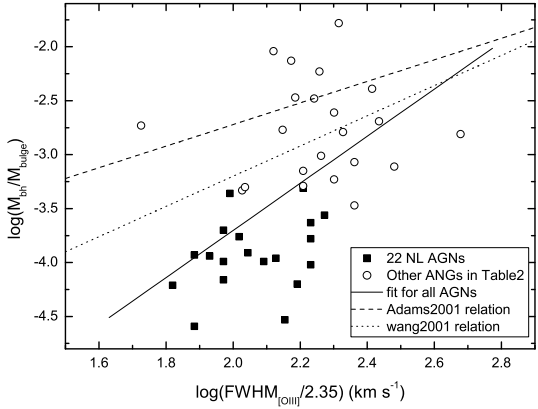}
\end{center}
\caption{MBH/Bulge mass ratio versus the stellar velocity
dispersion derived from FWHM of [O ${\rm \small III}$]. The solid
squares denote NL AGNs and the open circles denote BL quasars. The
thick solid line is the best fit to all objects with the available
FWHM of [O ${\rm \small III}$] (table 1 and table 2). The dashed
line is the theoretical line predicted by Adams et al. (2001) and
the dotted line is that from Wang et al. (2000).}
\end{figure}

Nelson and Whittle (1995) found that the stellar velocity
dispersion in the host galaxy can be converted from the [O {\rm
\small III}] FWHM in AGNs by $\sigma= \rm FWHM_{[O {\rm \small
III}]} /2.35$. Nelson (2001) has shown that the relation between
MBH mass and the bulge velocity dispersion derived from the FWHM
of [O {\rm \small III}] in AGNs is in good agreement with the
$M_{\rm bh}$ -- $\sigma$ relation defined by nearby inactive
galaxies. Wang \& Lu (2001) also show that it is the same for
NLS1s. We here use FWHM of the narrow line [O {\rm \small III}] as
a representation of the bulge velocity dispersion. The available
FWHM of [O ${\rm \small III}$] and $M_{\rm bh}/M_{\rm bulge}$,
except our 22 NL AHGNs, are listed in table 2. In figure 2 we plot
the MBH/Bulge mass ratio vs. the velocity dispersion for all
available data. We find that there is a moderately strong
correlation between them with $R=0.55$ ($P=2.25\times 10^{-4}$).
NGC 6814 is excluded in our fit because of its departure too much
from the trend. The MBH mass of NGC 6814 may be overestimated from
the overestimation of FWHM of H$\beta$ in Wandel (2002) (FWHM of
H$\beta$ in Wandel (2002) is 5500 $km~s^{-1}$ while that in Loar
(2001) is 3950 $km~s^{-1}$). Its low width of [O ${\rm \small
III}$] is also near to the resolving power. The fit line is
\begin{equation}
{\rm log}(M_{\rm bh}/M_{\rm bulge}) = (2.18\pm 0.54) {\rm log}
({\rm FWHM}_{\rm [O {\rm \small III}]}/(2.35 km~ s^{-1} )) -
(8.07\pm1.17).
\end{equation}
In figure 2 we also plot the theoretical lines from Adams et al.
(2001) and Wang et al. (2000), which are slightly different from
our fit line, but the trend that the MBH/Bulge mass ratio
increases with the bulge velocity dispersion is the same. We will
clarify it in the discussion section.

We also find that there is a weak correlation between bulge masses
and the FWHM of [O ${\rm \small III}$] for all of the available
data. The fit line is ${\rm log}(M_{\rm bulge}/\Msolar) =
(7.86\pm1.02) + (1.35\pm0.47) {\rm log} [{\rm FWHM}_{[{\rm O} {\rm
\small III}]}/(2.35 km~ s^{-1} )]$, with $R$ = 0.42 and $P$ =
0.006.

\subsection{The Eddington Ratio}

In this subsection, we calculated the ratio of the bolometric
luminosity, $L_{bol}$, to the Eddington luminosity, $L_{\rm Edd}$.
$L_{\rm bol}$ is usually calculated by $L_{\rm bol}=10 \lambda
L_{\rm 5100}$, where $L_{\rm 5100}$ is the monochromatic
luminosity at 5100 $\rm{\AA}$ (Wandel et al. 1999). Here, we adopt
the bolometric luminosity from Woo and Urry (2002), which was
taken from the spectral energy distribution (SED). The result of
the calculated Eddington ratio is presented in table 1. In figure
3 we plot the central MBH mass against $L_{\rm bol}/L_{\rm Edd}$
for our 22 NL AGNs and the best-fitting line for 72 AGNs by McLure
and Dunlop (2002). It is noteworthy that $L_{bol}/L_{Edd}$ for the
NL AGNs departs from the main trend determined by the line by
McLure and Dunlop (2002).

\begin{figure} \begin{center}
\FigureFile(80mm,80mm){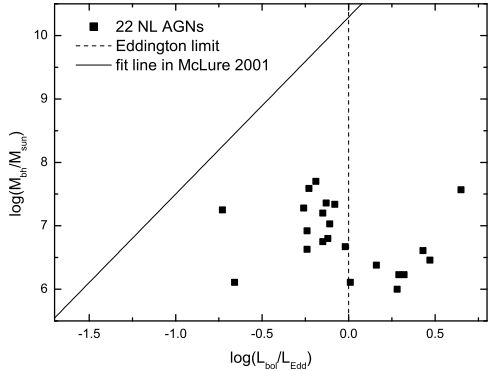} \end{center} \caption{MBH mass
versus bolometric luminosity as a fraction of the Eddington limit.
22 NL AGNs are shown as filled squares. The location of the
Eddington limit is shown by the vertical dash line. The best-fit
relation for 72 objects found by McLure \& Dunlop (2001) is shown
by the solid line. }
\end{figure}

\section{Discussion}
If the relation of the MBH mass to the bulge Luminosity is given
by $M_{\rm bh}\propto L_{\rm bulge}^{\alpha}$ ,the mass-to-light
ratio for the bulge is parameterized as $M_{\rm bulge}\propto
L_{\rm bulge}^{\beta}$, and the MBH/Bulge mass ratio is given by
$M_{\rm bh}\propto M_{\rm bulge}^{\gamma}$, we will find
$\gamma=\alpha/\beta$. Some authors give the relation between the
MBH mass and the bulge absolute $V$ magnitude with $M_{\rm bh}
\propto (M^{\rm bulge}_{V})^{\delta}$ , $\alpha=2.5\delta$
(equation 4). The value of $\beta$ that is commonly adopted is
1.18 (Magorrian 1998) or 1.31 (Jorgensen et al. 1996). Our result
and other authors' results for $\alpha, \beta, \gamma$ are list in
table 3. For the same MBH masses (figure 1), the NL AGNs have
larger bulge masses compared to the other BL AGNs and inactive
galaxies. NL AGNs are special, and should be dealt with separately
in a study of the MBH/Bulge relation.

\begin{table*}

\caption{The MBH/Bulge relation, where $M_{\rm bh}\propto L_{\rm
bulge}^{\alpha}$, $M_{\rm bulgr}\propto L_{\rm bulge}^{\beta}$,
$M_{\rm bh}\propto M_{\rm bulge}^{\gamma}$. }
\begin{center}
\begin{tabular}{llcccccc}
\hline \hline
Sample & Type& $N$ & $\alpha$ & $\beta$ & $\gamma$ & $R$ & ${\rm log} (M_{\rm bh}/M_{\rm bulge})$\\
(1)&(2)&(3)&(4)&(5)&(6)&(7)&(8)\\

\hline
Our NLS1s &NL AGNs&22&$1.90\pm0.70$&$1.18$&$1.61\pm0.59$&0.74&$-3.90\pm0.27$\\
Wandel 2002&NL AGNs&~9&$0.99\pm0.13$&1.18&$0.84\pm0.22$&0.82&$-3.85\pm0.29$\\
Wandel 2002 &BL AGNs &46&$0.90\pm0.11$&$1.18$&$0.76\pm0.09$&$0.78$&$-2.81\pm0.45$\\
McLure 2002 &AGNs&72&$1.15\pm0.08$&1.31&$0.88\pm0.06$&0.77&$-2.90\pm0.45$\\
Laor 2001&AGNs&24&$1.60\pm0.25$&$1.18$&$1.36\pm0.21$&$0.80$&- \\
Kormendy 2001$^{\dagger}$&Inactive&35&$0.96\pm0.13$&1.18&$0.81\pm0.10$&0.80&$-2.77\pm0.50$\\
Mathur 2001&NLS1s&15&-&-&-&-&$-4.33\pm0.47$\\

\hline
\end{tabular}
\end{center}
$\ast$ Col. 1 gives the sample. Col. 2 gives the class of the
objects. Col. 3 gives the number of objects in the sample. Col.
4-6 give the $\alpha$, $\beta$, $\gamma$. Col. 7 is the
correlation coefficient ($R$). Col. 8 gives the mean MBH/Bulge
mass ratio and the standard deviation. \\$\dagger$ Exclude NGC
4486B and NGC 5845 for their larger uncertainty in the MBH mass
(Wandel 2002).

\end{table*}

Although we obtained the MBH/Bulge mass ratios for 22 NL AGNs, we
should noticed that there are some uncertainties in the estimation
of the MBH/Bulge mass ratios. There are mainly several opinions
concerning the origin about the narrow width of H$\beta$ in NLS1s.
One is the small inclinations in NLS1s (figure 1 in McLure, Dunlop
2002; Bian, Zhao 2002); the second is the long distance of the
BLRs emitting line of H$\beta$ in NLS1s; the third is their higher
value of $L/L_{\rm Edd}$ because of their low central black hole
masses. The second option is more plausible considering the other
properties in NLS1s (Turner et al. 2002). The uncertainties in the
$B$ magnitude, continua, and the empirical equation 1 would lead
to an uncertainty of about 0.5 index in the MBH mass estimation
(Wang, Lu 2001).

The errors in the calculated bulge masses are mainly related to
the calculation of the bulge magnitude and the mass-light relation
for the bulge. The bulge luminosity obtained by a bulge/disk
decomposition of the galaxy images tends to be systematically
lower than that from the empirical formula for the bulge/total
ratio, depending on the Hubble type (Simien, de Vauculours 1986;
Wandel 2002). Wandel (2002) found a bulge luminosity correction
based on the width line of $H\beta$, which is derived from 15
Seyfert 1 galaxies common to the Wandel et al. (1999) sample and
the McLure and Dunlop (2001) sample. We don't use this bulge
luminosity correction because we find it larger than the value of
the bulge luminosity for the NL AGNs. Accurate values of the bulge
luminosity for NL AGNs is necessary in a study of the MBH/Bulge
relation in NL AGNs. In the mass -- light relation, $M_{\rm
bulge}\propto L_{\rm bulge}^{\beta}$, $\beta$ is usually adopted
as 1.18 since it is was determined through stellar dynamics
(Magorrian 1998). However, McLure and Dunlop (2001) assumed the
relation $M_{\rm bulge} \propto L_{\rm bulge}^ {1.31}$, which is
from the Gunn-r fundamental plane study (Jorgensen 1996). In this
paper we adopt $\beta=1.18$.

In figure 2, we find there is a correlation between the MBH/Bulge
mass ratio to the available velocity dispersion (from the FWHM of
[O ${\rm \small III}$]) for 22 NL AGNs and 22 BL AGNs, $M_{\rm
bh}/M_{\rm bulge}\propto \sigma^{2.18\pm0.54}$. We notice that the
relation is mainly due to the smaller MBH/Bulge mass ratio and the
smaller velocity dispersion for the NL AGNs. This relation can be
expected from the relation between the MBH mass and the bulge mass
and the relation between MBH mass and stellar velocity dispersion.
Our result gives $M_{\rm bulge}\propto M_{\rm bh}^{0.6}$ and
$M_{\rm bh}/M_{\rm bulge}\propto M_{bh}^{0.4}$. If $M_{\rm
bh}\propto \sigma^{a}$, then $M_{\rm bh}/M_{\rm bulge}\propto
\sigma^{1.92}$ [$a = 4.80$, Ferrarese et al. 2001], $M_{\rm
bh}/M_{\rm bulge}\propto \sigma^{1.50}$ [a  =3.75, Gebhardt et al.
2000b]. We suggested the nonlinear MBH/Bulge relation (Laor 2001).
This relation is consistent with some theoretical work (Wang et
al. 2000; Adams et al. 2001). We can't distinguish these two
models for their idealization.

Mathur (2000) has proposed that NLS1s are likely to be active
galaxies in an early stage of evolution. The mean MBH/Bulge mass
ratio of NLS1s will be significantly smaller than that of BL AGNs
and normal galaxies. A scenario of MBH growth is his preferred
interpretation. The accretion process determines the MBH mass
(Haehnelt et al. 1998). The Salpeter time for the growth of MBH,
i.e. the $e$-folding time, is $t=4\times 10^{7}(L_{\rm Edd}/L)
~\eta_{0.1}$ yr, where $\eta_{0.1}$ is the radiative efficiency
normalized to 0.1. Let us assume the calculated MBH masses to be
the initial MBH masses. The MBH would grow to a ``final'' Seyfert
mass, which is estimated from the MBH/Bulge mass ratio in BL AGNs
and the bulge masses. We adopt the MBH/Bulge mass ratio in BL AGNs
is 0.0012 (McLure, Dunlop 2002). The ``final'' Seyfert mass is
$0.0012M_{bulge}$. The growth time for NLS1s to a ``final''
Seyfert galaxy is $t_{\rm s}= {\rm log}_{e}(0.0012M_{\rm
bulge}/M_{\rm bh}) 4\times 10^{7}(L_{\rm Edd}/L_{\rm bol})
~\eta_{0.1}$ yr. Our calculated growth times of 22 NLS1s are
listed in Table 1. Since the accretion rate decreases with time,
the growth time is the lower limit. The mean growth time is
$(1.29\pm0.24)\times 10^{8}$ yr, which is close to the upper
limit, $4.5\times 10^{8} \eta_{0.1}$ yr calculated for $L_{\rm
bol}/L_{\rm Edd}=1$ (Haehnelt et al. 1998; Mathur et al. 2001).

Another interpretation of the nonlinear MBH/Bulge relation is that
it is also possible that that NL AGNs occur in low-$M_{\rm bulge}$
galaxies, and that in such galaxies $M_{\rm bh}/M_{\rm bulge}$ is
lower than in galaxies with a higher $M_{\rm bulge}$ if we
consider that NL AGNs already have their ``final'' $\rm M_{\rm
bh}/M_{\rm bulge}$. More information of the bulge about NL AGNs is
needed to clarify the black hole -- bulge relation in NL AGNs.

\section{Conclusion}
New MBH/Bulge mass ratios were calculated for a sample of 22 NL
AGNs using the FWHM of H$\beta$, nuclear $B$ magnitude and the
bulge absolute $B$ band magnitude. We obtained the mean MBH/Bulge
mass ratio and the MBH/Bulge relation. The main conclusions can be
summarized as follows:
\begin{itemize}
\item{ The mean of $M_{\rm bh}/M_{\rm bulge}$ for 22 NL AGNs is $-3.9\pm
0.07$, which is lower by one order of magnitude compared to that
of BL AGNs.}
\item{A correlation is found between the bulge
mass and the MBH mass for 22 NL AGNs (the correlation coefficient
is $R=0.74$), $M_{\rm bulge}\propto M_{\rm bh}^{0.62\pm0.13}$,
which is higher compared to that for BL AGNs. We suggest the
nonlinear MBH/Bulge relation. A correlation is found between the
MBH/Bulge mass ratio and the velocity dispersion converted from
the FWHM of [O ${\rm \small III}$] for 22 NL AGNs and 22 BL AGNs,
which is consistent with some recent theoretical studies. }
\item{A scenario of MBH growth for NL AGNs is one of
our interpretations of the nonlinear MBH/Bulge relation. The mean
MBH growth time for NLS1s to a ``final'' Seyfert galaxy is
$(1.29\pm0.24)\times 10^{8}$ yr. Another interpretation of the
nonlinear MBH/Bulge relation is also possible, that NL AGNs occur
in low $M_{\rm bulge}$ galaxies if we consider that NL AGNs
already have their ``final'' $M_{\rm bh}/M_{\rm bulge}$.}

\end{itemize}

\section*{Acknowledgements}
We thank the anonymous referee for the valuable comments. We thank
the financial support from Chinese Natural Science Foundation
under contract 10273007.


\begin{thebibliography}{}
\bibitem[]{} Adams, F. C., Graff, D. S., \& Richstone, D. O. 2001, ApJ,
551, L31
\bibitem[]{} Bahcall, J. N., Kirhakos, S., Saxe, D. H., \& Schneider, D. P. 1997, ApJ, 479, 462
\bibitem[]{} Bian, W., \& Zhao, Y.  2002, A\&A, 395, 465
\bibitem[]{} Blandford, R. D. 1999, in ASP Conf. Ser. 182, Galaxy
Dynamics, ed. D. R. Merritt, J.A.Shellwood, \& M. Valluri (San
Francisco:ASP)  p87

\bibitem[]{} Fabian, A. C. 1999, MNRAS, 308, L39
\bibitem[]{} Ferrarese, L., Pogge, R. W., Peterson, B. M., Merritt, D.,
Wandel, A., Joseph, C. L. 2001, ApJ, 555, L79

\bibitem[]{} Ferrarese, L., \& Merritt, D.  2000, ApJ, 539, L9
\bibitem[]{} Gebhardt, K., et al. 2000a, ApJ, 539, L13
\bibitem[]{} Gebhardt, K., et al. 2000b, ApJ, 543, L5
\bibitem[]{} Haehnelt, M. G., Natarajan, P., \& Rees M. J. 1998, MNRAS, 300,
817
\bibitem[]{} Ho, L. C. 1999, In ``Observational Evidence for Black
Holes in the Universe'', ed. S.K. Chakrabarti (Dordrecht: Kluwer)
p157
\bibitem[]{} Jorgensen, I., Franx, M., \& Kj$\ae$rgaard, P. 1996,
MNRAS, 280, 167
\bibitem[]{} Kaspi, S., Smith, P.S., Netzer, H., Maoz,
D., Jannuzi, B.T., Giveon, U. 2000, ApJ, 533, 631
\bibitem[]{} Kauffmann, G., \& Haehnelt, M. 2000, MNRAS, 311, 576
\bibitem[]{} Kormendy, J., \& Gebhardt, K. 2001, Proc. 20th
Texas Symposium, ed. H.Martel \& J.C.Wheeler (Austin:AIP) p363
\bibitem[]{} Kuraskiewicz J. et al. 2000, ApJ, 542, 692
\bibitem[]{} Laor, A. 1998, ApJ, 505, L83
\bibitem[]{} Laor, A. 2001, ApJ, 553, 677
\bibitem[]{} MacKenty, J. W. 1990, ApJS, 72, 231
\bibitem[]{} Magorrian, J., et al. 1998, AJ, 115, 2285
\bibitem[]{} Malkan, M. A., Gorjian, V., \& Tam, R. 1998, ApJS, 117, 25
\bibitem[]{} Marziani, Sulentic, J. W., Dultzin-Hacyan, D., Calvani, M. \& Moles, M. 1996, ApJS, 104, 37
\bibitem[]{} Mathur, S. 2000, MNRAS, 314, L17
\bibitem[]{} Mathur, S., Kuraszkiewicz, J., \& Czerney, B. 2001, New
Astronomy, 6,  321
\bibitem[]{} Merritt, D., \& Ferrarese, L. 2001, MNRAS, 320, L30
\bibitem[]{} McLure, R. J., \& Dunlop, J. S. 2001, MNRAS, 327, 199

\bibitem[]{} McLure, R. J., \& Dunlop, J. S. 2002, MNRAS, 331, 795
\bibitem[]{} Nelson, C. H.  2001, ApJ, 544, L91
\bibitem[]{} Nelson, C. H., \& Whittle, M. 1995, ApJS, 99, 67
\bibitem[]{} Silk, J., \& Rees, M. J. 1998, A\&A, 331, L1
\bibitem[]{} Simien, F., \& de Vaucouleurs, G., 1986, ApJ, 302, 564
\bibitem[]{} Turner, T. J., et al. 2002, ApJ, 568, 120
\bibitem[]{} Veron-Cetty, M.-P., Veron, P., \& Goncalves, A. 2001, A\&A,
372, 730
\bibitem[]{} Wandel, A., Peterson, B.M., \& Malkan, M.A. 1999, ApJ,
526,579
\bibitem[]{} Wandel, A. 2002, ApJ, 565,  762

\bibitem[]{} Wang, T. \& Lu, Y. 2001, A\&A, 377,52

\bibitem[]{} Wang, Y. P., Biermann, P.L., \& Wandel, A. 2000, A\&A, 361,
550

\bibitem[]{} Whittle, M. 1992, ApJS, 79, 49
\bibitem[]{} Woo, J. H., \& Urry, C. M. 2002, ApJ, 579, 530
\end{thebibliography}
\end{document}